\documentclass[floats,prd,aps,amssymb,nofootinbib,twocolumn]{revtex4}
\usepackage{amssymb}
\usepackage{graphics}
\usepackage{amsmath}
\usepackage{amsfonts}
\usepackage{bm}
\usepackage[]{latexsym}
\usepackage{cellspace}
\usepackage{mathtools}
\usepackage{amstext}
\usepackage{stmaryrd}
\usepackage{stackrel}
\usepackage{graphics}
\usepackage{graphicx}
\usepackage{esint}
\usepackage[utf8]{inputenc}
\usepackage{blindtext}
\usepackage{float}
\restylefloat{table}
\usepackage{booktabs}
\usepackage{enumitem} 
\usepackage{amssymb,amsbsy,epsfig,graphicx}
\usepackage{color}
\usepackage{etoolbox} 
\usepackage{lipsum} 
\usepackage{eufrak}
\usepackage{verbatim}
\usepackage{bm}
\usepackage{hyperref}
\usepackage[capitalize]{cleveref}
\hypersetup{pdfstartview=FitV,colorlinks=true,linkcolor=midblue,citecolor=midblue,filecolor=midblue,urlcolor=midblue}

\definecolor{midblue}{rgb}{0,0,0.5}

\newcommand{\mpl}{m_{\rm p}}

\newcommand{\be}{\begin{equation}}\newcommand{\ee}{\end{equation}}
\newcommand{\bea}{\begin{eqnarray}}\newcommand{\eea}{\end{eqnarray}}
\newcommand{\brr}{\begin{array}}\newcommand{\err}{\end{array}}
\newcommand{\bit}{\begin{itemize}}\newcommand{\eit}{\end{itemize}}
\newcommand{\ben}{\begin{enumerate}}\newcommand{\een}{\end{enumerate}}

\newcommand{\ba}{\begin{array}}
\newcommand{\ea}{\end{array}}

\begin{document}
	
\title{Phenomenology of GUP stars}

\author{Luca Buoninfante,\footnote{buoninfante.l.aa@m.titech.ac.jp}$^{\hspace{0.3mm}1}$ Gaetano Lambiase,\footnote{lambiase@sa.infn.it}$^{\hspace{0.3mm}2,3}$ Giuseppe Gaetano Luciano\footnote{gluciano@sa.infn.it}$^{\hspace{0.3mm}2,3}$ and Luciano Petruzziello\footnote{lpetruzziello@na.infn.it}$^{\hspace{0.3mm}2,4}$ }
 
\affiliation
{\vspace{1mm}$^1$Department of Physics, Tokyo Institute of Technology, Tokyo 152-8551, Japan,
\\ 
\vspace{0mm}
$^2$INFN, Sezione di Napoli, Gruppo collegato di Salerno, Fisciano (SA) 84084, Italy,
\\
\vspace{0mm}
$^3$Dipartimento di Fisica, Universit\`a di Salerno, Fisciano (SA) 84084, Italy,
\\
\vspace{0mm}
$^4$Dipartimento di Ingegneria, Universit\`a di Salerno, Fisciano (SA) 84084, Italy.}

  \def\be{\begin{equation}}
\def\ee{\end{equation}}
\def\al{\alpha}
\def\bea{\begin{eqnarray}}
\def\eea{\end{eqnarray}}

\begin{abstract}
We study quantum corrections at the horizon scale of a black hole induced by a Generalized Uncertainty Principle (GUP) with a quadratic term in the momentum. The interplay between quantum mechanics and gravity manifests itself into a non-zero uncertainty in the location of the black hole radius, which turns out to be larger than the usual Schwarzschild radius. We interpret such an effect as a correction which makes the horizon disappear, as it happens in other models of quantum black holes already considered in literature. We name this kind of horizonless compact objects {\it GUP stars}.  We also investigate some phenomenological aspects in the astrophysical context of binary systems and gravitational wave emission by discussing Love numbers, quasi-normal modes and echoes, and studying their behavior as functions of the GUP deformation parameter. Finally, we preliminarily explore the possibility to constrain such a parameter with future astrophysical experiments.
\end{abstract}

 \vskip -1.0 truecm
\maketitle

\section{Introduction}

Einstein's general relativity (GR) turns out to be the best description of the gravitational interaction so far. Its predictions have been tested to very high precision at large distances and late times \cite{-C.-M.}; just to mention the most recent experimental confirmations, we can think of the detection of gravitational waves (GWs) from a binary black hole merger~\cite{GW} and the capture of the first black hole image~\cite{Akiyama:2019cqa}. However, despite its great success, there are still conceptual problems which are plaguing the theory and no satisfactory solution has been found up today. Indeed, GR admits black hole solutions which suffer from curvature singularity and possess a horizon, and both these features cause troubles when short-distances are considered~\cite{Hawking} and/or quantization in curved spacetime is implemented~\cite{Giddings:2006sj,Unruh:2017uaw,Mathur:2009hf}. To be more precise, we have not seen any black hole experimentally, but only dark objects which behave very similarly to them, thus allowing for the existence of many alternative descriptions still not excluded by observations.

In order for a theory of gravity to be consistent, no singularity or paradox should arise. Recent investigations have shown that a resolution to the information loss paradox~\cite{Unruh:2017uaw,Mathur:2009hf} can be found by assuming that the effective radius of the astrophysical object is larger than the corresponding horizon radius, which becomes a feasible scenario when quantum corrections at the horizon scale are taken into account. For instance, two very well-known examples in this direction are represented by the fuzzball paradigm~\cite{Mathur:2005zp,Guo:2017jmi}, which has been proposed in string theory, and the gravastar~\cite{Mazur:2001fv,Mazur:2004fk,Beltracchi:2018ait,Beltracchi:2019tsu}.

Horizonless compact objects  have been intensely studied in the last years~\cite{Cardoso:2016oxy,cardoso-2016b,Cardoso:2017cqb,Carballo,wang,Holdom:2016nek,Ren:2019jft,Buoninfante:2018xiw,Oshita:2019sat,Wang:2019rcf,Saraswat:2019npa,Buoninfante:2020tfb} (see also Ref.~\cite{Cardoso:2019rvt} and therein for a detailed review on this topic).  Simpler models can also be constructed in terms of boson fields~\cite{Kaup:1968zz,Ruffini:1969qy,Lambiase:2017adh,Brito:2015pxa} without any reference to quantum corrections. However, the compactness turns out to be smaller in this context.

The main idea underpinning the aforementioned models is that the spacetime around a compact object can be well-described in terms of the Schwarzschild metric only up to some radius $R$ in Schwarzschild coordinates\footnote{In the following we adopt the mostly positive signature convention $\eta={\rm diag}(-+++)$ and work with natural units $\hbar=c=1$. In such units, the Newton constant is $G=1/M_p^2,$ with $M_p$ being the Planck mass, whereas the Planck length is $L_p=1/M_p.$}, i.e. (we consider the static case)
\begin{equation}
\begin{array}{rl}
ds^2=&\displaystyle -f(r)dt^2+\frac{dr^2}{f(r)}+r^2d\Omega^2\,,\\[3mm]
f(r)=&\displaystyle 1-\frac{R_H}{r},\quad {\rm for}\quad r\geq  R,
\end{array}
\end{equation}
where
\begin{equation}
R=R_H(1+\epsilon)\,,\quad \epsilon >0\,,
\end{equation}
with $R_H=2GM$ being the Schwarzschild radius. On the other hand, for $r<R$ the metric has to be modified, otherwise it might happen that no notion of geometry can be defined at all within the framework of quantized spacetime. It is believed that quantum gravity would modify the black hole radius through Planckian corrections, $R\sim R_H+L_p,$ so that the dimensionless parameter $\epsilon$ has to be very small, $\epsilon \ll 1$. Note that the standard black hole configuration can be smoothly recovered in the limit $\epsilon\rightarrow 0$.

One of the main features of horizonless compact objects is that, because of the absence of any horizon, there is a non-zero probability for an incoming wave to hit the surface of the object and then bounce back, or to enter inside the object and come out after some time. Therefore, modes can be trapped between the potential barriers located at the photon sphere and at the surface, thus generating periodic {\it echoes} which would add corrections to the GW signal of a black hole merger~\cite{Cardoso:2019rvt}. The generation of echoes provides a net observable signature which could discriminate between black holes and horizonless compact objects. We remark that it is not yet clear how to form such kind of ultra-compact objects (i.e., objects with compactness $\epsilon\ll 1$) and how their stability can be efficiently probed~\cite{Cardoso:2019rvt}. Therefore, the question is still open and surely deserves further investigation.

From an observational point of view, there have already been claims in favor of the observation of echoes in some of the GW event detection~\cite{GW}, but more work is still needed for a final conclusion to be reached~\cite{Abedi:2016hgu,Conklin:2017lwb,Abedi:2018npz,Ashton:2016xff,Westerweck:2017hus}. 
For the sake of completeness, let us stress that, although gravitational wave experiments are very useful to probe the horizon scale, the Event Horizon Telescope Collaboration~\cite{Akiyama:2019cqa} might not be able to reach those distances from the would-be horizon, as it mainly focuses on slightly larger scales closer to the photon sphere ($\sim 3GM$).

In this paper we aim to study quantum corrections to the black hole horizon induced by the Generalized Uncertainty Principle (GUP), which is expected to be one of the main ingredients in a quantum theory of gravity. By assuming a GUP with only a quadratic term in the momentum, we show that the ensuing corrections naturally imply the existence of quantum astrophysical objects, the radius of which is slightly larger than $R_H$. We do not interpret this effect as having black holes with a larger horizon, but we rather argue that such quantum corrections give rise to a new kind of horizonless objects, which we name {\it GUP stars}.

The plan is as follow. In Sec.~\ref{gup-BH-sec}, we briefly review the concept of uncertainty relations and subsequently introduce the notion of GUP star. In Sec.~\ref{pheno-sec}, we investigate some phenomenological aspects of GUP stars, such as Love numbers, quasi-normal modes, echoes and the absorption coefficient. Sec.~\ref{conclus-sec} is devoted to summary and conclusions.

\section{GUP stars}\label{gup-BH-sec}

The foundations of quantum mechanics rely on the Heisenberg Uncertainty 
Principle (HUP)~\cite{aHeisenberg:1927zz}
\begin{equation}
\delta x\,\delta p\geq \frac{1}{2}\,,\label{hup}
\end{equation}
which can be derived from the commutation relation
\begin{equation}
[\hat{x},\,\hat{p}]= i\,,
\label{comm-hup}
\end{equation}
where $\hat x$ and $\hat p$ are the position and momentum operators, respectively, and $\delta x$ and $\delta p$ the corresponding uncertainties. Therefore, quantum mechanics predicts that no measurement can determine position and momentum simultaneously with maximum precision, which also means that the phase space $x$-$p$ is discrete. 

Clearly, Eq.~\eqref{hup} as it stands does not say anything about the existence of a minimal length scale below which physics can no longer be probed. Such a scale inevitably appears if one also takes into account gravity effects, whose interplay with quantum mechanics limits the resolution of space and time~\cite{Maggiore:1993zu}.
Along this direction, several studies appeared in the past~\cite{Snyder:1946qz,Yang:1947ud,Karolyhazy:1966zz,Amati:1987wq,Gross:1987kza,Amati:1988tn,Konishi:1989wk,Maggiore:1993kv,Scardigli:1995qd,FS,Capozziello:1999wx,Bojowald:2011jd,Scardigli:2003kr,Adler:1999bu,Adler,Casadio:2009jc,Casadio:2013aua,scard,Paliathanasis:2015cza,Scardigli:2016pjs,VAF,Lambiase:2017adh,Scardigli:2018jlm,Luciano2019,Blasone:2019wad,BlasScard,Iorio:2019wtn}, which are mainly based on a 
proper generalization
of the HUP to the so-called Generalized Uncertainty Principle (GUP). One of the 
most commonly adopted form of GUP is
\begin{equation}
\delta x\,\delta p\ge \frac{1}{2}+2\hspace{0.3mm}\beta\hspace{0.3mm}\frac{\delta p^2}{M_p^2}\,,
\label{gup}
\end{equation}
where the {\it deformation parameter} $\beta$ is shown to be of order unity
in several approaches to quantum gravity, just like in string theory~\cite{Amati:1987wq,Amati:1988tn}, and in the context of Caianiello's theory of maximal acceleration~\cite{Capozziello:1999wx,Luciano2019}. Here, we assume $\beta>0,$ but we mention that scenarios with a negative value of the deformation parameter have also been investigated in literature~\cite{Jizba:2009qf,Ong:2018zqn,Kanazawa:2019llj,Buoninfante:2019fwr}. 

Furthermore, one can demonstrate that, for mirror-symmetric states 
(i.e. $\left\langle \hat{p}\right\rangle =0$), the inequality in Eq.~\eqref{gup} can be
obtained from the non-vanishing commutator 
\begin{equation}
[\hat{x},\,\hat{p}]= i\left[1+\beta\left(\frac{\hat{p}^2}{M_p^2}\right)\right].
\label{comm-gup}
\end{equation}
Let us emphasize that in the case of HEP we can describe quantum objects whose wavelength is inversely proportional to the mass, namely the De Broglie formula $\lambda\sim 1/M$ is valid. Instead, when both quantum mechanical and gravitational effects are not negligible, the GUP also induces a generalization of the De Broglie relation which becomes $\lambda\sim 1/M+M/M_p^2.$

We now apply the GUP framework to the study of black holes and astrophysical (ultra-)compact objects, which we assume to be intrinsically quantum in nature due to quantum mechanical effects taking place at the horizon scale \cite{Oshita:2019sat,Wang:2019rcf}. It is evident that because of their macroscopic structure ($M\gg M_p$), also gravitational effects are not negligible, and therefore a description in terms of GUP could be required. In particular, we assume that the mass-radius relation for such a compact object is given by the generalized De Broglie formula, consistently with Eq.~\eqref{gup}:
\begin{equation}
R =\frac{1}{2M}\,+\,2\hspace{0.3mm}\beta\hspace{0.3mm}\frac{M}{M_p^2}\,\,\, \Leftrightarrow\,\,\, R = \beta\hspace{0.1mm}R_H\left(1+\frac{1}{4\beta}\frac{M_p^2}{M^2}\right).
\label{ineq-BH}
\end{equation}
We can notice that, if $\beta=1$, the above relation takes the form
\begin{equation}
R= R_H\left(1+\frac{1}{4}\frac{M_p^2}{M^2}\right),
\label{ineq-BH-2}
\end{equation}
which means that the size of the object is {\it larger} than the Schwarzschild radius by an additional term which scales as the inverse of the mass squared. Note that this correction is intrinsically quantum: indeed, in the semi-classical limit $M/M_p\rightarrow \infty$, one recovers $R=R_H$, as it should be.

A comment is now in order. This kind of GUP corrections in the black hole context have already been analyzed in Refs.~\cite{Carr:2015nqa,Mureika:2018gxl}, where it is assumed that the effective radius $R$ is a new larger horizon. Therefore, the resulting gravitational object is still considered as a black hole, but of larger size. In this regard, we point out that our approach is different, as it relies on a completely novel point of view, which can be depicted as follows: in the semi-classical theory, the Schwarzschild black hole is a metric solution of Einstein's field equations describing the classical geometry of spacetime, and the concept of horizon is intrinsically geometric. 
It is also worth mentioning that, by taking into account the quantum back-reaction of matter fields on the background, one might already expect corrections to the black hole radius~\cite{Mottola:2010gp}. However, also in this case such a correction would be described in geometric terms, i.e. as an effective quantum corrected metric. Here, we want to stress that there is no reason why pure quantum corrections should be regarded a priori as a modification of the geometry. In fact, it is more reasonable to conceive them as fluctuations on the top of the geometrical background~\cite{Buoninfante:2020tfb}. In this way, the geometric notion of horizon can only emerge in the semi-classical limit, while it does not appear in a fully quantum scenario. 

Based on the outlined picture, in what follows we assume that GUP corrections give rise to quantum objects, the effective radius of which is always slightly larger than the corresponding Schwarzschild radius due to effects appearing at the horizon scale. Quantum mechanically speaking, this implies that a black hole is not really black.

Note that Eq.~\eqref{ineq-BH-2} has been obtained for $\beta=1$. However, we can be even more general and assume that the deformation parameter is not strictly equal to one, i.e.
\begin{equation}
\beta=1+\gamma\,,\quad\,\, \gamma\geq 0\,,
\label{gamma-deform}
\end{equation}
so that Eq.~\eqref{ineq-BH-2} becomes
\begin{equation}
R=R_H(1+\epsilon)\,,\,\,\quad\epsilon\equiv \gamma+\frac{1}{4}\frac{M_p^2}{M^2}\,.
\label{eff-radius}
\end{equation}
We call such objects {\it GUP stars}. Note that the semi-classical limit is now recovered when
\begin{equation}
\gamma\rightarrow 0\quad {\rm and}\quad \frac{M}{M_p}\rightarrow \infty \,.
\label{semi-class-limit}
\end{equation}
%


\section{Phenomenological aspects}\label{pheno-sec}

In this Section, we investigate some phenomenological aspects of GUP stars by analyzing crucial properties related to both the physics of GW and of the accretion disk; namely, we discuss both gravitational and electro-magnetic signals.

A physical quantity which turns out to be very useful when studying  
the physics of horizonless compact objects is the parameter $\mu$
which quantifies their {\it compactness} and is defined by 
\begin{equation}
\mu \equiv \displaystyle  1-\frac{R_H}{R}=\frac{\epsilon}{1+\epsilon}\,.
\label{compactness}
\end{equation}
It follows that the smaller $\mu$, the more compact the object (clearly, for Schwarzschild black holes, which are the most compact objects allowed in the theory, one has $\mu=0$). 

Now, for a GUP star, Eq.~\eqref{compactness} reads
\begin{equation}
\mu=\displaystyle \frac{\gamma+\frac{1}{4}\frac{M_p^2}{M^2}}{1+\gamma+\frac{1}{4}\frac{M_p^2}{M^2}}\,.
\label{compactness-gup}
\end{equation}
From the above relation, it is evident that the compactness coincides with the parameter $\epsilon$ in the $\epsilon \ll 1$ limit, that is
\begin{equation}
\epsilon \ll 1\,\,\,\, \Rightarrow\,\,\,\, \mu\simeq \epsilon =\gamma+\frac{1}{4}\frac{M_p^2}{M^2}\,,
\label{compactness-gup<<1}
\end{equation}
which is consistently satisfied when $\gamma, M_p/M\ll 1.$ 

In what follows, we shall work in the limit~\eqref{compactness-gup<<1} in which the compactness $\mu$ and the parameter $\epsilon$ are interchangeable, and write down our results in terms of $\epsilon$ only.

Before turning to a more quantitative
discussion of phenomenological aspects
of GUP stars, we stress that  
corrections to classical predictions
on the physics of black hole horizons are expected 
from more standard considerations of 
quantum field theory in curved space as well. 
In this framework one has that state--dependent
corrections to the stress-energy tensor should depend 
on the Schwarzschild radius as $\langle {T^a}_{b}\rangle\sim (\hbar /R^4_H)\left(1-R_H/r\right)^{-2}$~\cite{Mottola:2010gp}. Therefore, the effects of this stress tensor would become
relevant and affect the classical background geometry
for compactness values $\epsilon\sim L_p/R_H\sim M_p/M$. 
Nevertheless, as remarked in the previous Section, 
such corrections would have an intrinsic geometric description, 
i.e., they would be ascribed to an effective quantum
corrected metric rather than fluctuations of the background. 
We leave a more detailed comparative analysis 
of these effects and the ones induced by the GUP
for future investigation.

\subsection{Love numbers}\label{love-sec}

Let us consider two GUP stars of mass $M_1$ and $M_2,$ respectively, composing a binary system of total mass $M=M_1+M_2$ and ratio $q=M_1/M_2\geq 1$ (and with zero spin), which will end up in a new GUP star after merging. The spiral dynamics (before the merger) can be well-described in terms of Post-Newtonian (PN) perturbation theory, the expansion parameters of which are the weak gravitational field and the slow-velocity ($v^2\ll 1$). The main features are the energy and momentum loss of the binary system caused by multiple moments and tidal effects~\cite{Blanchet:2013haa}.

All dynamics is included in the produced  gravitational wave-form, which in Fourier space reads~\cite{Blanchet:2013haa}
\begin{equation}
\tilde{h}(f)=\mathcal{A}(f)\hspace{0.5mm}e^{i\left(\Psi_{\rm PP}+\Psi_{\rm TH}+\Psi_{\rm TD}\right)}\,,
\label{fourier-wave-form}
\end{equation}
where $f$ and $\mathcal{A}(f)$ are the GW frequency and amplitude, $\Psi_{\rm PP}$ takes into account point-particle interaction effects and $\Psi_{\rm TH}$ and $\Psi_{\rm TD}$ are related to tidal heating and tidal deformability, respectively. 

In Refs.~\cite{Yagi:2015hda,Uchikata:2016qku,Pani:2015tga}, it was shown that  the multipole moments approach those of a Schwarzschild black hole in the high compactness limit $\epsilon \ll 1$. Therefore, the phases $\Psi_{\rm TH}$ and $\Psi_{\rm TD}$ are more suitable to discriminate between the presence or absence of the horizon. The tidal heating takes into account the flux of energy that is absorbed by the object. In the case of a Schwarzschild black hole, it is maximal due to the presence of a horizon, while for a horizonless compact object it decreases, and for a perfectly reflecting surface one has $\Psi_{\rm TH}=0$~\cite{Maselli:2017cmm}. Since in the case of perfect reflection we can make analytic estimations for several quantities and better discuss phenomenology, we will mainly focus on the only remaining discriminator, i.e. $\Psi_{\rm TD},$ which is always vanishing for a Schwarzschild black hole~\cite{Binnington:2009bb,Damour:2009vw,Fang:2005qq,Poisson:2014gka,Pani:2015hfa}.

In a binary system the $\Psi_{\rm TD}$ phase takes into account the response of one of the two objects to the external gravitational field of the other, and the effect can be parametrized in terms of the so-called Love numbers, which can be non-zero for a horizonless object~\cite{Hinderer:2007mb,Flanagan:2007ix,Porto:2016zng,Cardoso:2017cfl}. The non-zero leading term for the phase $\Psi_{\rm TD}$ in the PN expansion is given by~\cite{Maselli:2017cmm,Addazi:2018uhd}
\begin{equation}
\begin{array}{rl}
\Psi_{\rm TD}(f)=& \displaystyle -\Psi_N \frac{\Lambda}{6M^5}v^{10}\frac{(1+q)^2}{q} \,,\\[3mm]
\Lambda\equiv &\displaystyle \left(1+\frac{12}{q}\right)M_1^5k_1+\left(1+12q\right)M_2^5k_2\,,
\end{array}
\label{phase-TD}
\end{equation}
where $\Psi_N=\frac{3}{128v^5}\frac{M^2}{M_1M_2},$ $v=(2\pi f)^{1/3}$ is the orbital velocity and corresponds to the PN expansion parameter, while $k_1$ and $k_2$ are the Love numbers depending on the internal structure of the objects (for a Schwarzschild black hole $k_i^{\rm BH}=0$). Therefore, tidal deformability introduces a $5$PN correction ($1$PN corresponds to $\sim v^2$) to the gravitational wave-form. As remarked above, in the case of perfect reflection one can obtain an analytic estimation for Love numbers~\cite{Cardoso:2017cfl},
\begin{equation}
k_i\sim \left|{\rm log}\,\epsilon_i\right|^{-1}\,,\quad i=1,2\,,
\end{equation}
which for a GUP star reads
\begin{equation}
k_i\sim\left|{\rm log}\left(\gamma +\frac{M^2_p}{4M_i^2}\right)\right|^{-1}\,,\quad i=1,2\,.
\label{love-number}
\end{equation}
From this relation, we deduce that GUP corrections introduce extra effects which would definitely discriminate GUP stars from semi-classical black holes. 

The future LISA experiment can measure and constrain Love numbers of the order of $k\sim 0.02,$ and possibly also reach $k\sim 0.005$ with further precision improvements. Thus, with the current LISA performance, we would be able to constrain the $\epsilon$ parameter to $\epsilon\lesssim e^{-1/(0.02)}\simeq 2\times 10^{-22}$. For a GUP star, this would translate into the condition
\begin{equation}
\gamma\lesssim 2\times 10^{-22}\,,
\label{gamma-bound-love}
\end{equation}
which, in terms of the deformation parameter $\beta$, can be read as
\begin{equation}
\beta-1\lesssim 2\times 10^{-22} \,.
\label{beta-bound-love}
\end{equation}
Note that such a bound does not improve the existing constraints on $\beta$ (see, for instance, Ref.~\cite{Lambiase:2017adh} and therein for a summary on all existing experimental bounds). In fact, current experiments on GUP-induced effects do not make any {\it a priori} assumption on the order of magnitude of $\beta$.  By contrast, in deriving Eq.~\eqref{beta-bound-love}, we have already assumed that $\beta=1+\gamma\sim \mathcal{O}(1),$ i.e. $\gamma\ll 1.$ However, the bound in Eq.~\eqref{gamma-bound-love} makes sense if we are interested in small deviations of the deformation parameter from unity, and indeed future experiments can allow us to constrain the smallness of the deviation $\gamma.$

\subsection{Quasi-normal modes}\label{qnm-sec}

The merger of two compact objects like GUP stars gives rise to an isolated system undergoing a relaxation phase during which quasi-normal modes (QNMs) are produced. QNMs can be described in terms of scalar, vector and tensor perturbations of the metric that can all be combined in one single differential equation. In Fourier space, such an equation reads~\cite{Zerilli:1971wd,Berti:2009kk}
\begin{equation}
\frac{d^2\psi(z)}{dz^2}+\left[\omega^2-V(r(z))\right]\psi(z)=0\,,
\label{ode-qnms}
\end{equation}
where $\omega=\omega_{R}+i\omega_{I}$ is a complex frequency ($\omega_R>0$ and $\omega_I<0$) and $z$ is the tortoise coordinate
\begin{equation}
z=r+2GM\,{\rm log}\left(\frac{r}{2GM}-1\right),
\label{tortois}
\end{equation}
with $M$ now being the mass of the final GUP star. For $r\geq R$, the potential $V(r)$ reads
\begin{equation}
V(r)=f(r)\left(\frac{l(l+1)}{r^2}+\frac{2GM(1-s^2)}{r^3}\right),
\label{potenti}
\end{equation}
with $l\geq s$ being the angular momentum and $s=0,-1,-2$ the spin of the perturbation (note that the dependence on $z$ is implicit in $r=r(z)$). In the case of spin $s=-2$, the potential differs for polar and axial perturbations: indeed, for axial perturbations, it is given by Eq.~\eqref{potenti} and is called Regge-Wheeler potential, while for polar perturbations we have the so-called Zerilli potential~\cite{Zerilli:1971wd}.

In the black hole case~\cite{Chandrasekhar:1975zza}, the QNMs solution to Eq.~\eqref{ode-qnms} can be found by imposing ingoing boundary condition at the horizon $r=2GM$ or, equivalently, at $z=-\infty$. In so doing, one can show that the fundamental QNM for $s=-2$ (i.e. $l=2$) is equal to $GM\omega\simeq 0.373672-i\hspace{0.5mm}0.0889623\,.$ Remarkably, since the ingoing boundary conditions are the same for both polar and axial perturbations, it so happens that the frequency spectrum of a black hole is isospectral.

Conversely, in the case of horizonless compact objects, the QNMs are different depending on the compactness of the objects, and also the isospectrality condition is generally not satisfied, so that  we have two sets of frequencies for polar and axial perturbations. This is due to the fact that the values of the Regge-Wheeler and Zerilli potentials at the surface $R$ are different, while in the case of a Schwarzschild black hole they both vanish at $r=2GM.$
Thus, in the presence of a surface $R\geq 2GM$, the solution of Eq.~\eqref{ode-qnms} close to the surface is given by \cite{Cardoso:2019rvt}
\begin{equation}
\psi(z\simeq z(R))=A_{in}(\omega)e^{-i\omega z}+A_{out}(\omega)e^{i\omega z}\,,
\label{sol-clos-surface}
\end{equation}
where $A_{in}(\omega)$ and $A_{out}(\omega)$ represent ingoing and outgoing amplitudes, and are related to the absorption and reflection coefficients, respectively. For a Schwarzschild black hole, it is clear that $A^{\rm BH}_{out}=0$ due to the presence of a horizon.

In what follows we assume that the surface of the object is perfectly reflecting ($A_{in}=0$), so that we can make a direct comparison with the (totally absorbing) black hole case. Remarkably, for a mirror surface one can obtain an analytic estimation for the real and imaginary parts of the quasi-normal frequencies of a static horizonless object with $\epsilon\ll 1$ \cite{Maggio:2017ivp,Maggio:2018ivz}:
\begin{equation}
M\omega_R\sim M_p^2\left| {\rm log}\,\epsilon \right|^{-1}\,,\quad M\omega_{I}\sim -M_p^2\left|{\rm log}\,\epsilon\right|^{-(2l+3)}\,.
\label{QNF-estimat}
\end{equation}
For a GUP star, these relations translate into
\begin{equation}
\begin{array}{rl}
M\omega_R\sim &\displaystyle M_p^2\left| {\rm log}\left(\gamma+\frac{M_p^2}{4M^2}\right)\right|^{-1}\,,\\[3.5mm]
M\omega_{I}\sim & \displaystyle -M_p^2\left|{\rm log}\left(\gamma+\frac{M_p^2}{4M^2}\right)\right|^{-(2l+3)}\,,
\end{array}
\label{QNF-GUP}
\end{equation}
which consistently hold for $\gamma\ll 1$ and $M/M_p\gg 1.$ 

Hence, GUP corrections at the horizon scale would drastically change the dynamics after the merger, since the resulting astrophysical object will relax in a very different way from a Schwarzschild black hole.

\subsection{Echoes}\label{echoes-sec}

In the absence of any horizon, ingoing waves leaving the photon sphere can be reflected by the surface and trapped between the two potential barriers at $r\simeq R$ and $r\simeq 3GM,$ with the latter corresponding approximatively to the maximum of the potential in Eq.~(\ref{potenti}). As a consequence, some of the trapped waves can cross the photon sphere and become outgoing towards space-like infinity after bouncing off the surface. Such a phenomenon significantly modifies the late time behavior of the wave-form introducing periodic {\it echoes}, the amplitudes of which decrease in time,  while the early time dynamics soon after the merger turns out to be the same as in the black hole case (indeed, the first burst is determined by the black hole QNM \cite{Cardoso:2016oxy}). Let us also stress that, in the absence of a photon sphere, no echo would be generated as no trapping region would exist. Thus, a necessary condition to have echoes is $0<\epsilon <0.5,$ which already puts the upper bound $\gamma\lesssim 0.5.$

The most important scale is given by the period $T_{echo}$ for the roundtrip of a wave to go from the photon sphere to the surface and come back \cite{Cardoso:2016oxy,Cardoso:2019rvt}. This characteristic time is given by
\begin{equation}
\begin{array}{rl}
T_{echo}=&\displaystyle 2\int_{R_H(1+\epsilon)}^{\sim 3GM} \frac{dr}{f(r)} \\[6mm] 
=& 2GM-4GM\epsilon - 4GM{\rm log}(2\epsilon)\,, 
\end{array}
\label{echo-time}
\end{equation}
which for a GUP star reads
\begin{equation}
\!T_{echo}=\frac{2M}{M_p^2}\left(1\!-2\gamma\!-\!\frac{M_p^2}{2M^2}\right)\! -\! 4\frac{M}{M_p^2}{\rm log}\left(2\gamma+\frac{M_p^2}{2M^2}\right).  
\label{echo-time-GUP}
\end{equation}
In the $\epsilon\ll 1$ limit (i.e. for $\gamma\ll 1$ and $M/M_p\gg 1$), we then obtain
\begin{equation}
T_{echo}\sim - \frac{M}{M_p^2}{\rm log}\,\epsilon\sim  - \frac{M}{M_p^2}{\rm log}\left(\gamma+\frac{M_p^2}{4M^2}\right). 
\label{echo-time-small epsilon}
\end{equation}
From the above equation, it follows that GUP corrections can be in one-to-one correspondence with the late time behavior of the wave-form by dictating the rhythm over which the wave form dies off. 

Let us now observe that, for $\beta=1$ (i.e. $\gamma=0$), the echo time scale in Eq.~\eqref{echo-time-small epsilon} approximatively reads $T_{echo}\sim {\rm log}\left(M/M_p\right)$.
It is worthwhile emphasizing that this last expression for $T_{echo}$ tells us that the echo time scale is of the same order of the scrambling time as pointed out in Refs.~\cite{Abedi:2016hgu,Saraswat:2019npa}.

\subsection{Absorption coefficient}

So far, we have mainly worked in the case of a perfectly reflecting surface, i.e. with the boundary condition $A_{in}(\omega)=0$ (see the discussion below Eq.~\eqref{sol-clos-surface}). In this Subsection, we also include a non-zero absorption coefficient. Moreover, we study some phenomenological implications coming from the electro-magnetic wave sector in relation to the physics of the accretion disk of a supermassive object at the center of a galaxy, where we assume a GUP star to be located.

The first relevant ingredient is the solid angle $\Delta \Omega$ under which photons coming out of the compact object can escape from the surface of radius $R$ and reach spatial infinity. Of course, for a Schwarzschild black hole, this angle is zero because of the presence of the horizon. By contrast, for a horizonless object ($\epsilon\neq 0$) it is non-vanishing  \cite{Cardoso:2017njb,Abramowicz:2002vt} and for a GUP star reads
\begin{equation}
\begin{array}{rl}
\displaystyle \frac{\Delta\Omega}{2\pi}=&\hspace{-1mm}\displaystyle \frac{27}{8}\hspace{0.5mm}\epsilon+\mathcal{O}(\epsilon^2) \\[3mm]
\simeq &\hspace{-1mm}\displaystyle \frac{27}{8} \left(\gamma+\frac{M_p^2}{4M^2}\right).
\end{array}
\label{solid-angle}
\end{equation}
Therefore, the more compact the object is, the darker it appears. 

By following Ref.~\cite{Carballo}, let us now introduce the following observables describing the physics of the surface of astrophysical compact objects: 
\begin{itemize}
	\item $\kappa$ is the {\it absorption coefficient} and stands for the amount of absorbed energy;
	
	\item $\Gamma$ is the {\it elastic reflection coefficient} and measures the fraction of ingoing energy that bounces off the surface elastically;
	
	\item $\tilde{\Gamma}$ is known as {\it inelastic reflection coefficient} and takes into account the fraction of energy which crosses the surface going inside the compact object, and comes out after a certain amount of time, i.e. the amount of energy that is reflected inelastically.
\end{itemize}
Note that the following consistency relation must hold true: $\kappa+\Gamma+\tilde{\Gamma}=1$. This is clearly satisfied for a semi-classical black hole, for which $\kappa_{\rm BH}=1$, $\Gamma_{\rm BH}=\tilde{\Gamma}_{\rm BH}=0$.

In the absence of a horizon, the compact object can exchange energy with the accretion disk around it, namely there may be a non-vanishing outgoing flux of energy as described by the relation~\cite{Carballo}
\begin{equation}
\frac{\dot{E}}{\dot{M}_{\rm disk}}\simeq \frac{(1-\kappa-\Gamma)(1-\Gamma)\Delta\Omega/2\pi}{\kappa+(1-\kappa-\Gamma)\Delta\Omega/2\pi}\,,
\label{relation-absorp}
\end{equation}
where $\dot{M}_{\rm disk}$  is the amount of energy loss per unit time of the accretion disk  falling towards the compact object, while $\dot{E}$ stands for the amount of energy per unit of time emitted by the compact object. As expected, in the limit $\epsilon \rightarrow 0$ ($\Delta\Omega\rightarrow 0$) we recover $\dot{E}_{\rm BH}\rightarrow0.$

Astronomical observations of the accretion disk dynamics allow us to constrain the quantity in Eq.~\eqref{relation-absorp}, thus yielding the following upper bound~\cite{Broderick:2007ek,Narayan:2008bv,Broderick:2009ph,Carballo}:
\begin{equation}
\frac{(1-\kappa-\Gamma)(1-\Gamma)\Delta\Omega/2\pi}{\kappa+(1-\kappa-\Gamma)\Delta\Omega/2\pi}\lesssim \mathcal{O}(10^{-2}).
\label{uppper-bound-rel}
\end{equation}
This means that, if we know the numerical value of $\kappa$ and $\Gamma$, we can find an experimental bound on $\epsilon$ or, in other words, on $\gamma$ by using the inequality
\begin{equation}
\gamma\lesssim \frac{8}{27}\frac{\kappa\cdot \mathcal{O}(10^{-2})}{(1-\kappa-\Gamma)(1-\Gamma)}-\frac{M_p^2}{4M^2}\,.
\label{ineq-bound-2}
\end{equation}
For instance, if we assume that the GUP star is a very good absorber, by choosing $\kappa\simeq 0.9,$ $\Gamma\simeq 0.05$ and $\tilde{\Gamma}\simeq0.05,$ we obtain 
\begin{equation}
\gamma\lesssim \mathcal{O}\left(10^{-1}\!-\!10^{-2}\right),
\label{ineq-bound}
\end{equation}
where we have exploited the condition $M\gg M_p$. 
In terms of the deformation parameter $\beta$, the above relation 
can be read as
\begin{equation}
\beta-1\lesssim \mathcal{O}\left(10^{-1}\!-\!10^{-2}\right).
\label{beta-bound-em}
\end{equation}
As already discussed below Eq.~\eqref{beta-bound-love}, our constraints hold if we assume $\gamma$ to be a small deviation from $\beta=1$. Indeed, within our framework, Eq.~\eqref{relation-absorp} relies on Eq.~\eqref{solid-angle} which only holds for $\epsilon\ll 1,$ i.e. $\gamma\ll 1.$ With such an assumption, we can truly exploit future astrophysical observations to constrain small deviations of the deformation parameter from unity.

\section{Conclusions}\label{conclus-sec}
In this paper, we have investigated quantum corrections at the horizon scale of a Schwarzschild black hole induced by a Generalized Uncertainty Principle with a quadratic term in the momentum. In agreement with other general treatments of quantum black holes \cite{Abedi:2016hgu,Oshita:2019sat,Wang:2019rcf}, we have argued that GUP-induced effects make the horizon disappear, giving rise to a horizonless astrophysical object, the size of which is larger than the usual Schwarzschild radius. We named this kind of object GUP star.

We have addressed several phenomenological aspects of GUP stars, first in relation to the physics of the gravitational waves emitted during the merger of a binary system, and then to the dynamics of the accretion disk at the centre of a galaxy. Specifically, we have studied the tidal deformability by estimating Love numbers of a GUP star, the quasi-normal modes and the generation of echoes, which turns out to be one of the main distinctive features due to the absence of horizons. In order to make analytic estimations, all these quantities have been computed by assuming that the surface of the GUP star is perfectly absorbing. Subsequently, we have relaxed such an assumption, introducing a non-zero absorption coefficient. In this context, we have studied the energy exchange between a GUP star and its surrounding accretion disk at the centre of a galaxy.

Finally, we have explored the possibility to constrain the GUP deformation parameter. Based on the predictions of some models of string theory and assuming small deviations of $\beta$ from the unit value, i.e. $\beta=1+\gamma\sim \mathcal{O}(1)$, we have shown that future astrophysical experiments might allow us to put very strong constraints on the parameter $\gamma$ or, in other words, on $\beta-1$. More work is inevitably required along this direction.

\vspace{0.4cm}

\acknowledgments

L. B. acknowledges support from JSPS no. P19324 and KAKENHI Grant-in-Aid for Scientific Research no. JP19F19324.

\end{document}